\documentclass[amssymb,amsmath,aps,groupedaddress,reprint,superscriptaddress,showpacs]{revtex4-1}

\usepackage{graphicx}
\usepackage[caption=false]{subfig}
\usepackage{dcolumn}
\usepackage{bm}
\usepackage{amsmath}
\usepackage{physics}
\usepackage[utf8]{inputenc}
\usepackage[english]{babel}
\usepackage{color}
\usepackage{stackengine}

\begin{document}
\author{A. Ryan Kutayiah}
\affiliation{Department of Physics and Astronomy, Texas A\&M
University, College Station, TX, 77843 USA}
\title{Difference frequency generation of surface plasmon-polaritons in Landau quantized graphene}
\author{Mikhail Tokman}
\affiliation{Institute of Applied Physics, Russian Academy of Sciences}
\author{Yongrui Wang}
\affiliation{Department of Physics and Astronomy, Texas A\&M
University, College Station, TX, 77843 USA}
\author{Alexey Belyanin}
\affiliation{Department of Physics and Astronomy, Texas A\&M
University, College Station, TX, 77843 USA}

\date{\today}

\begin{abstract}

We develop a rigorous quantum-mechanical theory of the nonlinear optical process of difference frequency generation of surface plasmon-polaritons in Landau-quantized graphene. Although forbidden in the electric-dipole approximation,  the second-order susceptibility is surprisingly high, equivalent to the bulk magnitude above $10^{-3}$ m/V. We consider the graphene monolayer as a nonlinear optical component of a monolithic photonic chip with integrated pump fields.  The nonlinear power conversion efficiency of the order of tens $\mu$W/W$^2$ is predicted from structures of $10-100$ $\mu$m size. We investigate a variety of waveguide configurations to identify the optimal geometry for maximum efficiency.

\end{abstract}


\maketitle

\section{Introduction}

Many of the unique transport, thermal, and optical properties of graphene stem from the fact that its low-energy excitations are massless Dirac fermions \cite{CastroNeto2009}.  Among its numerous applications is the use of graphene as an optoelectronic and plasmonic material.  Graphene was shown to support highly-confined surface plasmon modes \cite{Griorenko2012,Bludov2013}; it has relatively long-lived plasmon-polariton modes due to large intrinsic carrier mobilities and doping tunability \cite{Rana2008,Jablan2009,Koppens2011},  excellent electro-optic tunability \cite{Cracium2011}, and  large third-order and second-order optical nonlinearity \cite{mikhailov2009,Hendry2010,Yao2012,yao2014,wang2016}.  The latter is surprising since graphene is a centrosymmetric medium for low-energy in-plane excitations. Therefore, its in-plane  second-order nonlinear response should be zero in the electric dipole approximation \cite{Boyd2008}. However, for obliquely incident or in-plane propagating electromagnetic (EM) fields, inversion symmetry is broken by nonzero wavevector components in the plane of graphene, and the second-order nonlinearity is nonzero and actually quite large \cite{Mikhailov2011,yao2014,wang2016,tokman2016,sipe2017}. It is enabled by effects of the spatial dispersion, or, in real space, by nonlocal effects beyond the electric dipole approximation. A particularly large value of $\chi^{(2)}$ equivalent to the bulk value of $ \sim 10^{-3}$ m/V per monolayer \cite{wang2016} is reached at low frequencies, for the processes of frequency down-conversion to the terahertz range such as difference frequency generation (DFG) \cite{yao2014,wang2016,Jamalpoor2017,Yao2017,Cao2018} or parametric down-conversion \cite{tokman2016}. 

A strong magnetic field transverse to the graphene layer splits the continuous conical electron dispersion into a discrete set of non-equidistant Landau levels (LLs) \cite{Goerbig2011}. The magnetic field does not break the inversion symmetry, so the DFG process remains forbidden in the electric dipole approximation. However, a strong magnetic field creates resonant transitions for all EM fields and enhances the electron density of states through the LL degeneracy. Both effects enhance optical nonlinearity \cite{Yao2012,Yao2013}.  Further enhancement of the nonlinear generation efficiency is possible when the DFG signal is frequency- and phase-matched to surface plasmon-polaritons in graphene. 
   
This work focuses on DFG in Landau-quantized graphene, particularly on the nonlinear generation of surface plasmon polaritons. In Section II we derive the dispersion equation for surface plasmon-polaritons in Landau-quantized graphene. In Section III We calculate the second-order nonlinear susceptibility and generated DFG signal power. For calculations of the Poynting flux of nonlinearly generated surface plasmon-polaritons, we focus on the monolithically integrated photonic chip geometry, including graphene as a nonlinear material and a dielectric waveguide or cavity with strong vertical confinement for the pump electromagnetic (EM) fields. We obtain analytic expressions for the DFG plasmon power and present its dependence on various parameters. We investigate a variety of waveguide configurations to identify the optimal geometry for maximum DFG efficiency. Our results can be easily extended to other (non-waveguide) geometries of the pump beams delivery and overlap.

\section{Dispersion of surface plasmon-polaritons in Landau-quantized graphene}

\begin{figure}[ht]

\subfloat[Graphene is located at the interface  $z=-d/2$ of dielectrics with dielectric constants $\epsilon_{2}$ and $\epsilon_{3}.$]{%
  \includegraphics[clip,width=0.5\columnwidth]{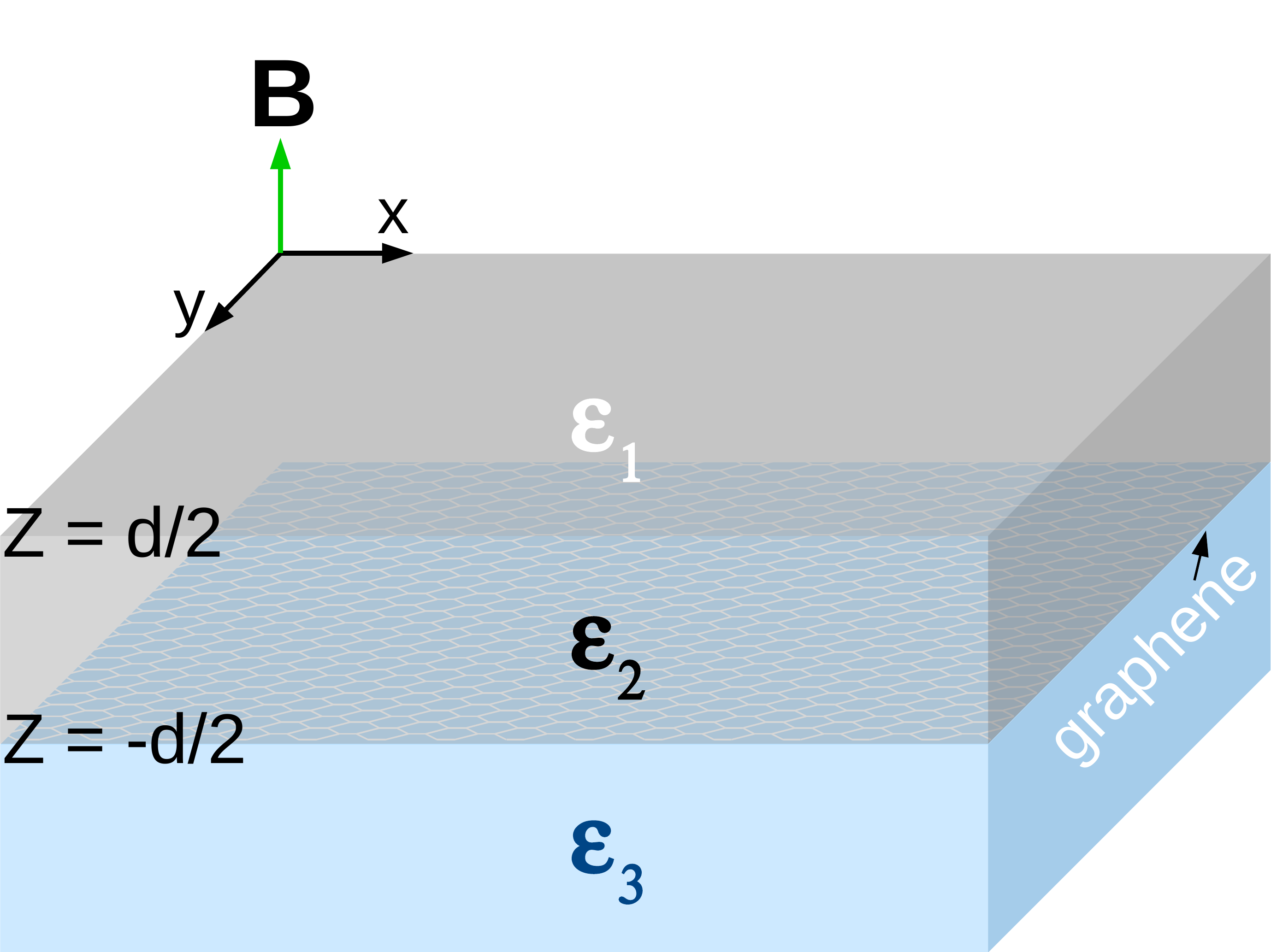}\label{onea}%
}
\subfloat[Graphene is located at the center of the waveguide core with dielectric constant $\epsilon_{2}$ in the $z=0$ plane.]{%
  \includegraphics[clip,width=0.5\columnwidth]{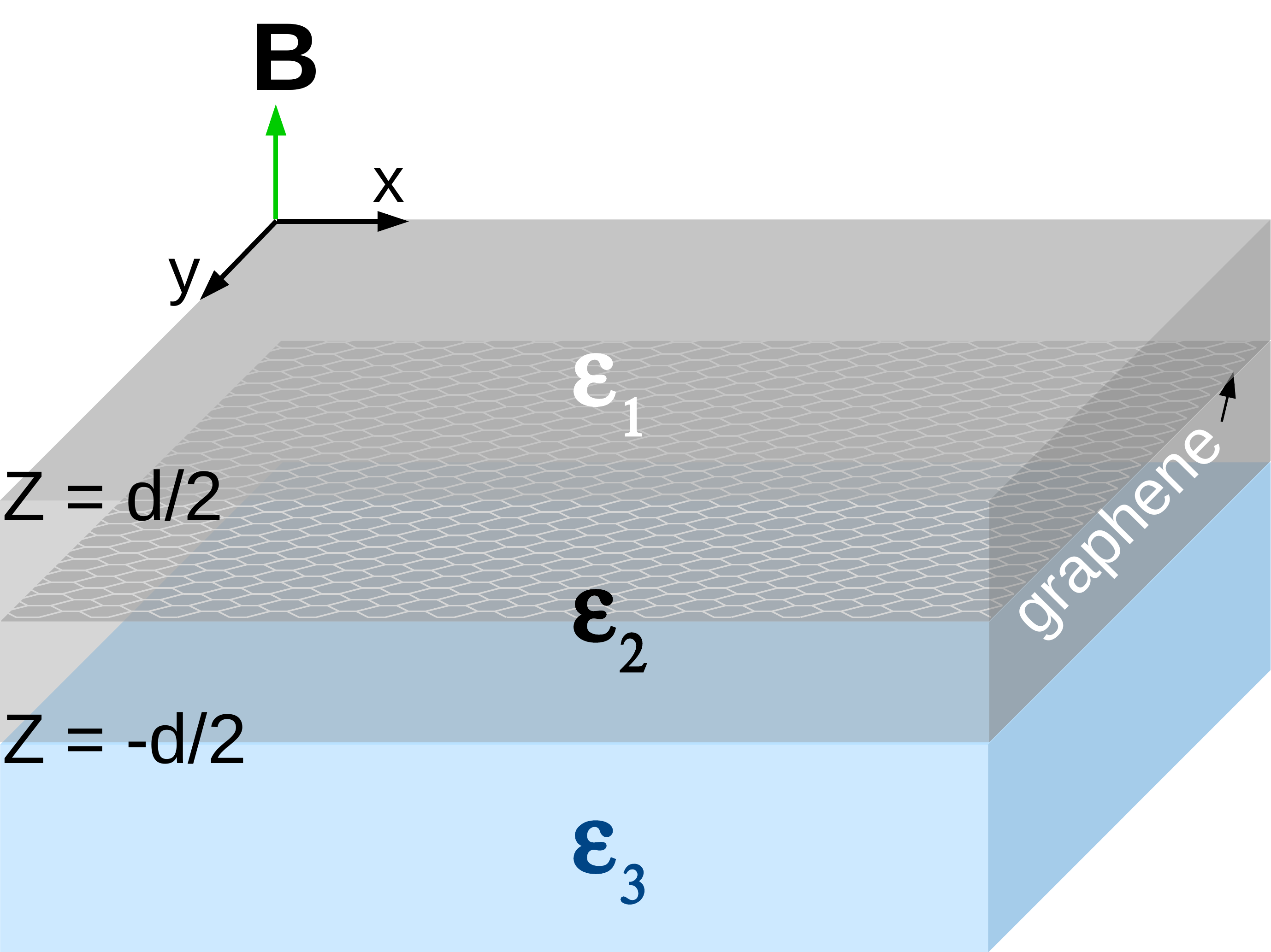}\label{oneb}%
}

\caption{A sketch of integrated waveguide geometries.   }

\end{figure}

We consider two possibilities for integrating  a monolayer of graphene of area $S$ into a dielectric waveguide or cavity;  see Fig.~\ref{onea} and Fig.~\ref{oneb}.  There is a uniform magnetic field in the $z$-direction $\bm{B}=\bm{e_z}B$. 
  
The dielectric constants of the waveguide layers $\epsilon_{1}$, $\epsilon_{2}$, and $\epsilon_{3}$ will be taken as air, GaAs, and AlAs (respectively) or air, Si, and SiO$_{2}$ (respectively) for numerical examples below. However, many other combinations of the cladding and core layer materials are possible with the same qualitative results.  The pump modes participating in the DFG are guided by the waveguide core $\epsilon_2$ and are counterpropagating in the $x$-direction to provide phase matching to surface plasmon-polaritons (SPPs) supported by graphene at the difference frequency. This arrangement ($\epsilon_{1}\neq\epsilon_{2}\neq\epsilon_{3}$) is what we refer to as the asymmetric waveguide.  The special case where $\epsilon_{1}=\epsilon_{3}$ (both are air) is what we will call the symmetric waveguide. 

Note that the effect of disorder in adjacent dielectric layers can be very detrimental for carrier mobility and optical transition linewidth in graphene. To avoid an excessive linewidth broadening, graphene should be encapsulated between hexagonal boron nitride (hBN) layers. We will neglect the thickness of dielectric hBN layers in the calculations of EM modes assuming that they are of nm thickness. They can be easily taken into account if needed.  

\subsection{Surface charge density for Landau-quantized graphene}

The surface charge density for graphene in a magnetic field is given by
\begin{equation}\label{surfchargeden}
\rho(\bm{r})= -e\sum_{\alpha,\beta}\rho_{\alpha\beta} \psi_{\beta}^{*}(\bm{r})\psi_{\alpha}(\bm{r})
\end{equation}
where $e$ is the elementary charge, $\rho_{\alpha\beta}$ is the density matrix, $\psi_{\alpha}(\bm{r})$ are the energy eigenstates for Landau-quantized graphene near the Dirac point as given in Appendix A, i.e. $\psi_{\alpha}(\bm{r})=\psi_{nk}(\bm{r})$.  They form a complete set and are orthonormal in the area $S$.  The vector $\bm{r}$ is in the graphene plane. The index $\alpha$ is a shorthand notation for electron quantum numbers $n,k$ in a magnetic field.  

Next we evaluate the spatial Fourier transform of the surface charge density,
\begin{equation}\label{fourier}
\rho(\bm{r})=\sum_{q}\rho_{q}e^{i\bm{q}\cdot\bm{r}}, \quad \rho_{q}=\frac{1}{S}\int d^{2}r\, e^{-i\bm{q}\cdot\bm{r}}\rho(\bm{r}),
\end{equation}   
where $\bm{q}$ is in the plane of graphene.  Substituting Eq.~(\ref{surfchargeden}) into the integral in Eq.~(\ref{fourier}) gives
\begin{equation}\label{discretesurfchargeden}
\rho_{q}=-\frac{e}{S}\sum_{\alpha,\beta}F_{\beta\alpha}(-\bm{q})\rho_{\alpha\beta}
\end{equation} 
where  $F_{\beta\alpha}(-\bm{q}) = \mel{\beta}{e^{-i\bm{q}\cdot\bm{r}}}{\alpha}$. Assuming that $\bm{q}$ is directed along $x$, we obtain 
\begin{align}
&F_{\beta\alpha}(-\bm{q})=\mel{n,k'}{e^{-iqx}}{m,k}\nonumber\\
&=\mel{n,k'}{e^{-iqx}}{m,k'+q}\delta_{k,k'+q}\equiv \tilde{F}_{nk'm}(-q)\delta_{k,k'+q}\label{effectivedipmom}.
\end{align} 
The matrix element $\tilde{F}_{nk'm}(-q)$ is calculated in Appendix B. 

One needs to solve the density matrix equation for $\rho_{\alpha\beta}$ to obtain the Fourier component of the surface charge density $\rho_{q}$.  We assume that the electric field of a graphene SPP in the plane of graphene is described by a scalar potential $\Phi(\bm{r},t) = Re[\Phi_{q}e^{i\bm{q}\cdot\bm{r}-i\omega_{q}t}]$. The density matrix equation is:
\begin{align}
\dot{\rho}_{\alpha\beta}&+\frac{i}{\hbar}(\mathcal{E}_{\alpha}-\mathcal{E}_{\beta})\rho_{\alpha\beta}+\rho_{\alpha\beta}\gamma_{\alpha\beta}\nonumber\\
=&-\frac{i}{\hbar}(e\Phi_{q}e^{i\bm{q}\cdot\bm{r}-i\omega_{q}t})_{\alpha\beta}(f_{\alpha}-f_{\beta})\label{densitymateq}\\
\implies &\rho_{\alpha\beta}(t)=\frac{-eF_{\alpha\beta}(\bm{q})(f_{\alpha}-f_{\beta})}{\hbar(\omega_{\alpha\beta}-\omega_{q}-i\gamma_{\alpha\beta})}\Phi_{q}e^{-i\omega_{q}t}.\label{densitymateqsol}
\end{align}
Here we used the rotating wave approximation, $\gamma_{\alpha\beta}$ is the phenomenological decay term for a transition between states $\ket{\alpha}$ and $\ket{\beta}$, $f_{\alpha}=\rho_{\alpha\alpha}$ is the occupation number of a given state. 

\subsection{Dispersion relation for graphene surface plasmon-polaritons }  

The dispersion relation for SPPs in the quasi-electrostatic regime $q>>\omega_{q}/c$ is obtained by  using Gauss' law in 2D, the solution of the Laplace equation in a uniform dielectric (see also  \cite{yao2014}), 
\begin{equation}
(\epsilon_{2}+\epsilon_{3})q\Phi_{q}=4\pi\rho_{q}, \label{gauss2D}
\end{equation}
and the relationship between the surface charge density and polarization, which is the definition of the surface linear susceptibility, 
\begin{equation}
\rho_{q}=-q^{2}\chi_{\|}(\omega_{q},\bm{q})\Phi_{q}. \label{continuityeqn}
\end{equation}
The above two equations yield the dispersion relation
\begin{equation}
 D(\omega_{q},\bm{q})=1+\frac{4\pi q}{\epsilon_{2}+\epsilon_{3}}\chi_{\|}(\omega_{q},\bm{q})=0. \label{dispersion}
\end{equation}
Using Eqs.~(\ref{discretesurfchargeden}), (\ref{densitymateqsol}), and (\ref{continuityeqn}) one arrives at the expression for the surface linear susceptibility,
\begin{equation}\label{linsusceptibility}
\chi_{\|}(\omega_{q},\bm{q})=-\frac{e^2}{Sq^{2}}\sum_{\alpha,\beta}\frac{(f_{\alpha}-f_{\beta})|F_{\alpha\beta}(\bm{q})|^{2}}{\mathcal{E}_{\alpha}-\mathcal{E}_{\beta}-\hbar\omega_{q}-i\hbar\gamma_{\alpha\beta}}.
\end{equation}
Inserting Eq.~(\ref{linsusceptibility}) into Eq.~(\ref{dispersion}), results in the dispersion relation for a SPP in Landau-quantized graphene  
\begin{equation}\label{fulldispersion}
D(\omega_{q},\bm{q})=1-\frac{4\pi e^{2}}{(\epsilon_{2}+\epsilon_{3})Sq}\sum_{\alpha,\beta}\frac{(f_{\alpha}-f_{\beta})|F_{\alpha\beta}(\bm{q})|^{2}}{\mathcal{E}_{\alpha}-\mathcal{E}_{\beta}-\hbar\omega_{q}-i\hbar\gamma_{\alpha\beta}}.
\end{equation}

To avoid cumbersome expressions we consider resonant three-wave mixing when both pump modes and the difference frequency signal are resonant to three cascaded inter-LL transitions and form a closed loop, as shown in Fig.~2. Then it is enough to consider  three Landau levels $\ket{-n}$, $\ket{n-1}$, and $\ket{n+1}$ which we relabel $\ket{1}$, $\ket{2}$, and $\ket{3}$, respectively.  The pump fields at frequencies $\omega_{1}$ and $\omega_{2}$ are coupled to electric-dipole allowed transitions $\ket{1}\rightarrow\ket{3}$ and $\ket{1}\rightarrow\ket{2}$ which obey the selection rules $\Delta |n|=\pm 1$. However, the difference-frequency transition $\ket{2}\rightarrow\ket{3}$, or $\ket{n-1} \rightarrow \ket{n+1}$, does not and is therefore electric-dipole forbidden. This is another manifestation of the fact that DFG is electric-dipole-forbidden in monolayer graphene. 

We assume that the Fermi level is somewhere between states $\ket{2}$ and $\ket{3}$ but separated by more than $k_{B}T$ from state $\ket{3}$, see Fig.~2.  The pump modes are TE-polarized and counterpropagating, in order to satisfy phase-matching conditions for a DFG of SPPs. Their frequencies are resonant with transitions $\ket{1}\rightarrow\ket{3}$ and $\ket{1}\rightarrow\ket{2}$, respectively, i.e. $\omega_{1}\approx \omega_{31}$ and $\omega_{2}\approx \omega_{21}$.

Using the states given above and the fact that $f_{3k}\approx 0$ the dispersion relation (\ref{fulldispersion}) becomes
\begin{equation}\label{altfulldispersion}
D(\omega_{q},\bm{q})=1+\frac{\omega_{o}(q)}{\omega_{32}-\omega_{q}-i\gamma_{32}}=0
\end{equation}
where
\begin{equation}\label{omegao}
\omega_{o}(q)=\frac{4\pi e^{2}(N_{F}/S)\xi(q)}{(\epsilon_{2}+\epsilon_{2})\hbar q},\quad \xi(q)=\frac{\sum_{k}|\tilde{F}_{3k2}(q)|^{2}}{\kappa}
\end{equation}
where $\kappa=2S/\pi l_{B}^{2}$ is the Landau level degeneracy, $N_{F}=f_{F}\kappa$ is the number of particles in a completely filled Landau level, and $f_{F}=f_{2k'}$.  It follows from Eq.~(\ref{altfulldispersion}) that
\begin{equation}\label{omegaq}
{\rm Re}[\omega_{q}]=\omega_{32}+\omega_{o}(q),\quad {\rm Im}[\omega_{q}] = - \gamma_{32}.
\end{equation}

\begin{figure}[ht]
\includegraphics[clip,width=0.9\columnwidth]{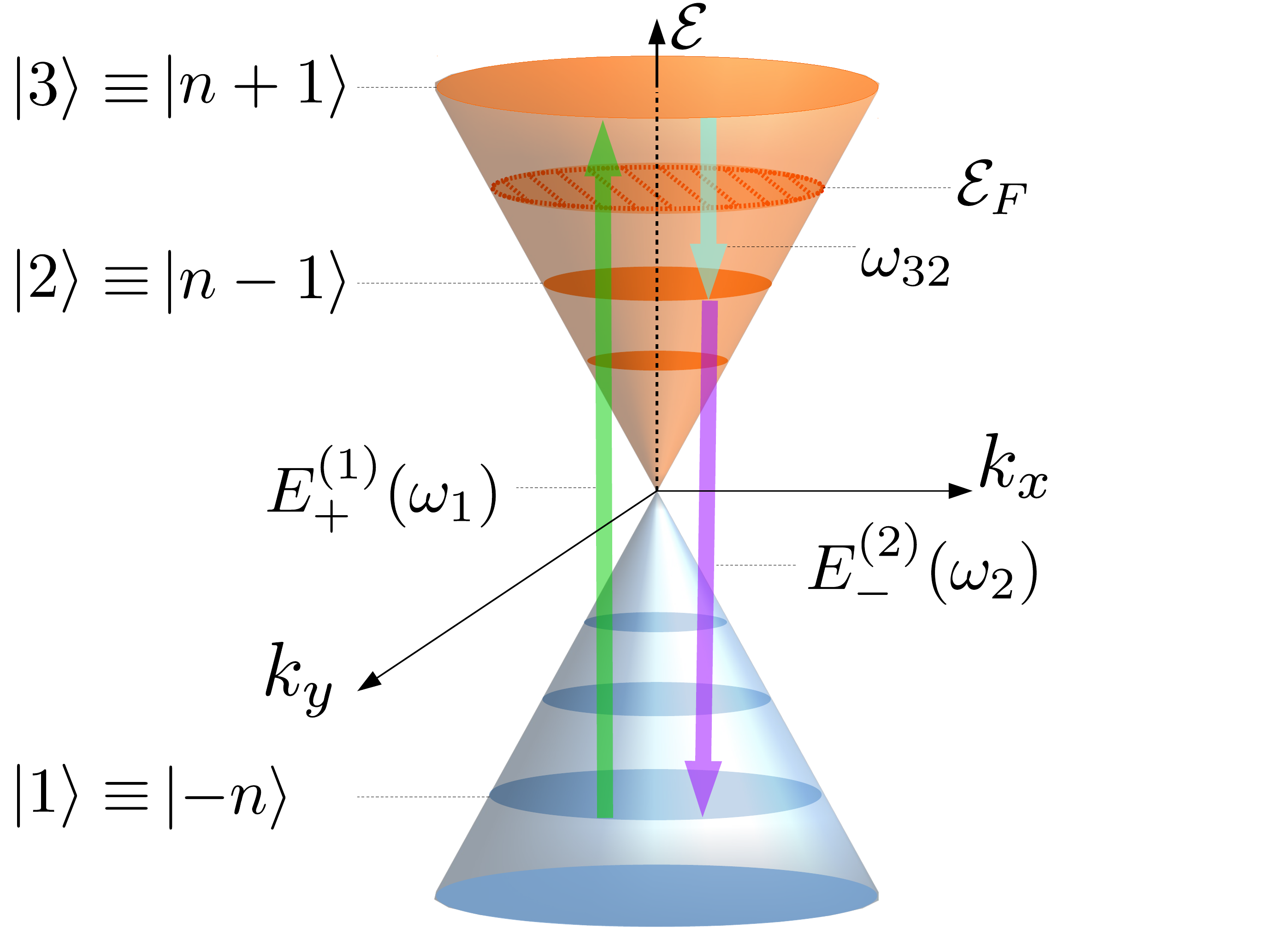}\label{landaulevelsaxes}
\caption{A sketch of Landau levels in graphene (not to scale) superimposed on the Dirac cone and the resonant DFG scheme.  The pump fields $E_{\pm}$ are coupled to electric-dipole allowed Landau level transitions. The difference frequency field is resonant to a dipole-forbidden transition.}
\end{figure}

For a dipole-forbidden transition $\ket{2}\rightarrow\ket{3}$  $|\tilde{F}_{3k2}(\bm{q})|^{2}\propto q^{a>2}$ when $q$ is small. For large $q$ the quantity $|\tilde{F}_{3k2}(\bm{q})|^{2}$ goes to zero too.  

\section{Difference-frequency generation of SPPs in Landau-quantized graphene}

\subsection{Nonlinear charge density and second-order susceptibility}

In the presence of the pump fields generating the polarization at the difference frequency, the surface charge density needs to be expanded to include nonlinear terms,
\begin{equation}
\label{expandeddensity} 
\rho_{q}=\rho_{q}^{l}+\rho_{q}^{nl}.
\end{equation}
Here we identify the linear part as the one linearly proportional to the electric field, $\rho_{q}^{l}=-q^{2}\chi_{\|}(\omega_{q},\bm{q})\phi_{q}$ (compare with Eq.~(\ref{continuityeqn})), where $\phi_{q}$ is the harmonic of the scalar potential of the SPP field and the nonlinear term as $\rho_{q}^{nl}$.   By inserting Eq.~(\ref{expandeddensity}) into Eq.~(\ref{gauss2D}) one can solve for $\phi_{q}$ in terms of the nonlinear part of the charge density $\rho_{q}^{nl}$, 
\begin{equation}\label{nonlinquasielectrostaticamp}
\phi_{q}=\frac{4\pi\rho_{q}^{nl}}{(\epsilon_{2}+\epsilon_{3})qD(\omega_{q},\bm{q})}.
\end{equation}

To derive the expression for the nonlinear charge density we express the Fourier component of the nonlinear surface charge density $\rho_{q}^{nl}$ in terms of its matrix elements as done in Eq.~(\ref{discretesurfchargeden}). Following \cite{tokman2013},   we obtain the equation for the density matrix element $\rho_{3k2(k-q)}$, which corresponds to the transition $\ket{3}\rightarrow\ket{2}$,
\begin{align}
\dot{\rho}_{3k2(k-q)}+i\omega_{32}\rho_{3k2(k-q)}+\gamma_{32}\rho_{3k2(k-q)}&\nonumber\\
=-i\frac{d^{*}_{21}E^{(2)*}_{-}(-d/2)e^{i\omega_{2}t}}{\hbar}\rho_{3k1(k-q_{1})}&\label{densmateq32}.
\end{align} 
Here and in all equations below the pump fields $E^{(1)*}_{+}, E^{(2)*}_{-}$ are taken on the graphene monolayer located at $z = -d/2$. Therefore, below we omit the argument $-d/2$ in the pump fields.  Furthermore, $q=q_{1}+q_{2}$, where $q_{1,2}$ are the projections of wavevectors of the optical fields on the graphene plane. 

 We see that the density matrix for the transition $\ket{3}\rightarrow\ket{2}$ depends on the linear perturbation of the matrix element for the transition $\ket{3}\rightarrow\ket{1}$.  There is no contribution from the density matrix for the transition $\ket{2}\rightarrow\ket{1}$ because states $\ket{1}$ and $\ket{2}$ are below the Fermi level and assumed fully occupied.  Consequently, the difference in their population is zero. We will assume that the pump fields are not strong enough to cause significant population transfer. 

The density matrix element for the $\ket{3}\rightarrow\ket{1}$ transition can be solved within the electric dipole approximation,  
\begin{align}
\dot{\rho}_{3k1(k-q_{1})}+i\omega_{31}\rho_{3k1(k-q_{1})}+\gamma_{31}\rho_{3k1(k-q_{1})}&\nonumber\\
=i\frac{d_{31}E_{+}^{(1)}e^{-i\omega_{1}t}}{\hbar}f_{F},&\label{densmateq31}\\
{\rm or} \; \rho_{3k1(k-q_{1})}(t)=\frac{e^{-i\omega_{1}t}f_{F}}{\omega_{31}-\omega_{1}-i\gamma_{31}}\frac{d_{31}E_{+}^{(1)}}{\hbar}&\label{densmateq31sol}.  
\end{align}
Inserting Eq.~(\ref{densmateq31sol}) into Eq.~(\ref{densmateq32}) yields:
\begin{align}
\rho_{3k2(k-q)}(t)=-\frac{e^{-i(\omega_{1}-\omega_{2})t}f_{F}}{(\omega_{32}-(\omega_{1}-\omega_{2})-i\gamma_{32})(\omega_{31}-\omega_{1}-i\gamma_{31})}&\nonumber\\
\times\frac{d_{31}d_{21}^{*}E_{+}^{(1)}E_{-}^{(2)*}}{\hbar^{2}}&\label{densmateq32sol}.
\end{align}
We are now equipped with almost all the pieces to express the amplitude of the SPP field in terms of the pump field amplitudes.  The last piece of information we need is the expression for $\rho_{q}^{nl}$ which is obtained from Eqs.~(\ref{discretesurfchargeden}), (\ref{effectivedipmom}), and (\ref{densmateq32sol}),
\begin{align}
\rho_{q}^{nl}(t)=\frac{(N_{F}/S)\zeta(q)e^{-i\omega_{d}t}}{(\omega_{32}-\omega_{d}-i\gamma_{32})(\omega_{31}-\omega_{1}-i\gamma_{31})}&\nonumber\\
\times\frac{ed_{31}d_{21}^{*}E_{+}^{(1)}E_{-}^{(2)*}}{\hbar^{2}}&\label{nonlindens},
\end{align}
where $\omega_{d}=\omega_{1}-\omega_{2}$, and $\zeta(q)=\sum_{k'}\tilde{F}_{2k'3}(-\bm{q})/\kappa$.  The matrix elements entering the expression for $\zeta(q)$ are evaluated in Appendix B. 

Note that the second-order nonlinear susceptibility $\chi^{(2)}$ can be extracted from Eq.~(\ref{nonlindens}) by using  $\rho_{q}^{nl}=-i\bm{q}\cdot\bm{P}^{nl}_{q}=-iq\chi^{(2)}E_{+}E_{-}^{*}$:
\begin{equation}\label{chi2}
\chi^{(2)}(\omega_{q},\bm{q})=\frac{i}{q}\frac{(N_{F}/S)\zeta(q)}{(\omega_{32}-\omega_{d}-i\gamma_{32})(\omega_{31}-\omega_{1}-i\gamma_{31})}\frac{ed_{31}d_{21}^{*}}{\hbar^{2}}.
\end{equation}
The magnitude of $\chi^{(2)}$ scales linearly with $q$. For a range of $q$ corresponding to DFG of THz plasmons by mid-infrared pumps, and for $B = 1$ T, $|\chi^{(2)}| \sim 2 \times 10^{-7}$ in CGS units. Just for the sake of comparison with nonlinear crystals, we can divide by graphene monolayer thickness to get the ``bulk'' magnitude of $|\chi^{(2)}_{3D}| \sim 3 \times 10^{-3}$ m/V, which is a very large number. Of course, the resulting DFG power efficiency depends on the magnitue of the surface (2D) $\chi^{(2)}$, as well as  the overlap of modes with graphene and the sample size.

Finally, the expression of the field amplitude of the SPP mode can be obtained by  substituting Eq.~(\ref{nonlindens}) into Eq.~(\ref{nonlinquasielectrostaticamp}),
\begin{align}
\phi_{q}=&\frac{4\pi}{(\epsilon_{2}+\epsilon_{3})qD(\omega_{q},\bm{q})}\nonumber\\
&\times\frac{ed_{31}d_{21}^{*}E_{+}^{(1)}E_{-}^{(2)*}(N_{F}/S)\zeta(q)}{\hbar^{2}(\omega_{32}-\omega_{d}-i\gamma_{32})(\omega_{31}-\omega_{1}-i\gamma_{31})}\label{ampofquasielectrostaticmodefull1}.
\end{align}
After making use of Eq.~(\ref{altfulldispersion}) and some straightforward manipulations one arrives at the final expression for the Fourier harmonic of the scalar potential of the SPP field:
\begin{align}
&\phi_{q}=\frac{4\pi e(N_{F}/S)\zeta(q)}{(\epsilon_{2}+\epsilon_{3})q}\nonumber\\
&\times \frac{(d_{31}d_{21}^{*}E_{+}^{(1)}E_{-}^{(2)*})/\hbar^{2}}{(\omega_{32}+\omega_{0}(q)-(\omega_{1}-\omega_{2})-i\gamma_{32})(\omega_{31}-\omega_{1}-\gamma_{31})}\label{ampofquasielectrostaticmodefull2}.
\end{align}

\subsection{Poynting Flux in a SPP Mode}

In the quasi-electrostatic approximation the time derivative of the magnetic field of the electromagnetic wave is negligible.  In order to calculate the Poynting flux of the transverse magnetic (TM) SPP mode we need to go beyond the quasi-electrostatic approximation. Using Maxwell's equations (see also  \cite{yao2014}), we derive all required components of the electric and magnetic fields starting from the tangential component of the electric field, that is, the field along the $x$-axis of the graphene monolayer:
\begin{equation}\label{parallelEfieldingraphene}
E_{xq}(z=-d/2)\equiv E^{o}_{xq}=-iq\phi_{q};
\end{equation} 
\begin{align}
E_{x}(x,z,t)&=E_{xq}^{o}e^{iqx-i\omega_{q}t}\label{parallelEfieldgraphenefull}
\begin{cases}
e^{-p_{2}(z+d/2)} & \quad z>-d/2\\
e^{+p_{3}(z+d/2)} & \quad z<-d/2
\end{cases},\\
E_{z}(x,z,t)&=\pm\frac{iq}{p_{2,3}}E_{x}(x,z,t),\\
B_{y}(x,z,t)&=\mp\frac{i\omega_{q}\epsilon_{2,3}}{cp_{2,3}}E_{z}(x,z,t),
\end{align}
where $p_{2,3}=\sqrt{q^{2}-\epsilon_{2,3}\omega^{2}_{q}/c^{2}}>0$ is the inverse confinement length in the $z$-direction. In $\pm$ or $\mp$ the top sign corresponds to $z>-d/2$ and the bottom sign corresponds to $z<-d/2$. 

The Poynting flux is then
\begin{align}\label{Poyntingflux}
\bm{S}&=\frac{c}{8\pi}(\bm{E}\times\bm{B}^{*})\\
&=\bm{e_x}\frac{\omega_{q}q^{3}}{8\pi}|\phi_{q}|^{2}
\begin{cases}
\frac{\epsilon_{2}}{p^{2}_{2}}e^{-2p_{2}(z+d/2)}, &\; z>-d/2\\
\frac{\epsilon_{3}}{p^{2}_{3}}e^{+2p_{3}(z+d/2)}, &\; z < -d/2
\end{cases}
\end{align}

To calculate the power in the SPP mode at the difference frequency, we integrate the Poynting flux Eq.~(\ref{Poyntingflux}) over the differential area $\bm{e_x}dydz\rightarrow\bm{e_x}L_{y}dz$, assuming that a graphene sample is uniform in the $y$-direction.  The power is then 
\begin{equation}\label{DFGpower}
P_{DFG}=\frac{L_{y}\omega_{q}q^{3}|\phi_{q}|^{2}}{16\pi}\left(\frac{\epsilon_{2}}{p^{3}_{2}}+\frac{\epsilon_{3}}{p^{3}_{3}}\right).
\end{equation}
In the approximation $q \gg \omega_{q}/c$ we can write $p_{2,3}\approx q$, $q^{3}(\epsilon_{2}/p_{2}^{3}+\epsilon_{3}/p_{3}^{3})\approx \epsilon_{2}+\epsilon_{3}$.  Using this approximation along with Eq.~(\ref{ampofquasielectrostaticmodefull2}) gives the final expression for the SPP power: 
\begin{align}
&P_{DFG}=\frac{\pi L_{y}(\omega_{1}-\omega_{2})}{\epsilon_{2}+\epsilon_{3}}\left[\frac{e(N_{F}/S)}{q}\right]^{2}\nonumber\\
&\times\left|\frac{(d_{31}d^{*}_{21}E^{(1)}_{+}E^{(2)*}_{-}/\hbar^{2})\zeta(q)}{(\omega_{32}+\omega_{o}(q)-(\omega_{1}-\omega_{2})-i\gamma_{32})(\omega_{31}-\omega_{1}-i\gamma_{31})}\right|^{2}.\label{DFGpowerfull}
\end{align}
This expression was derived for the graphene monolayer at the interface of the dielectric waveguide core and cladding.  Similar formulas can be obtained for any other location of graphene. 

\begin{figure}[h]
    \centering
    \includegraphics[width=0.8\columnwidth]{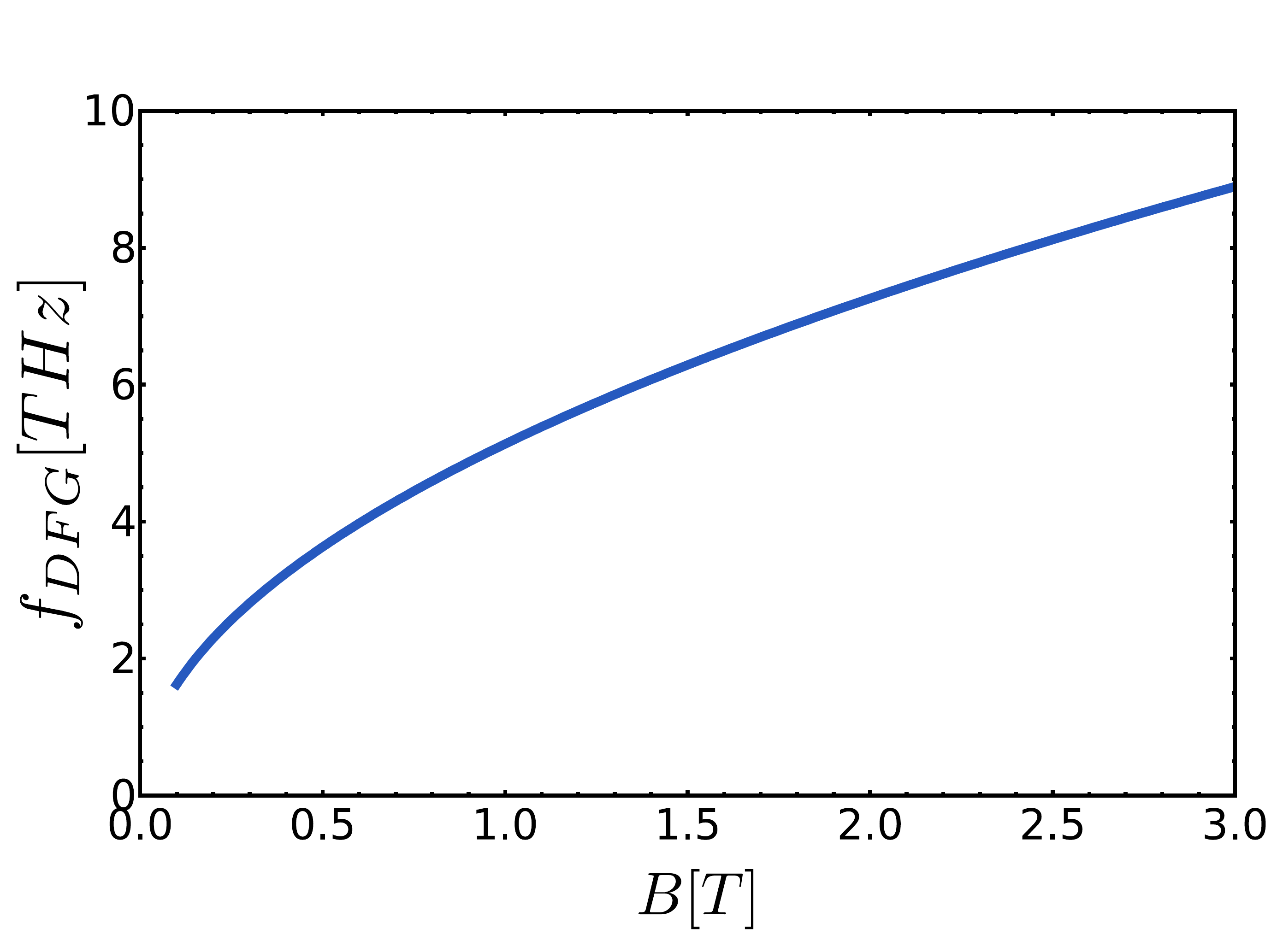}
    \caption{The DFG frequency resonant to the transition between Landau-level numbers 2 and 4 as a function of the magnetic field strength.}
	\label{fig3}
\end{figure}

\begin{figure}[ht]
\includegraphics[clip,width=\columnwidth]{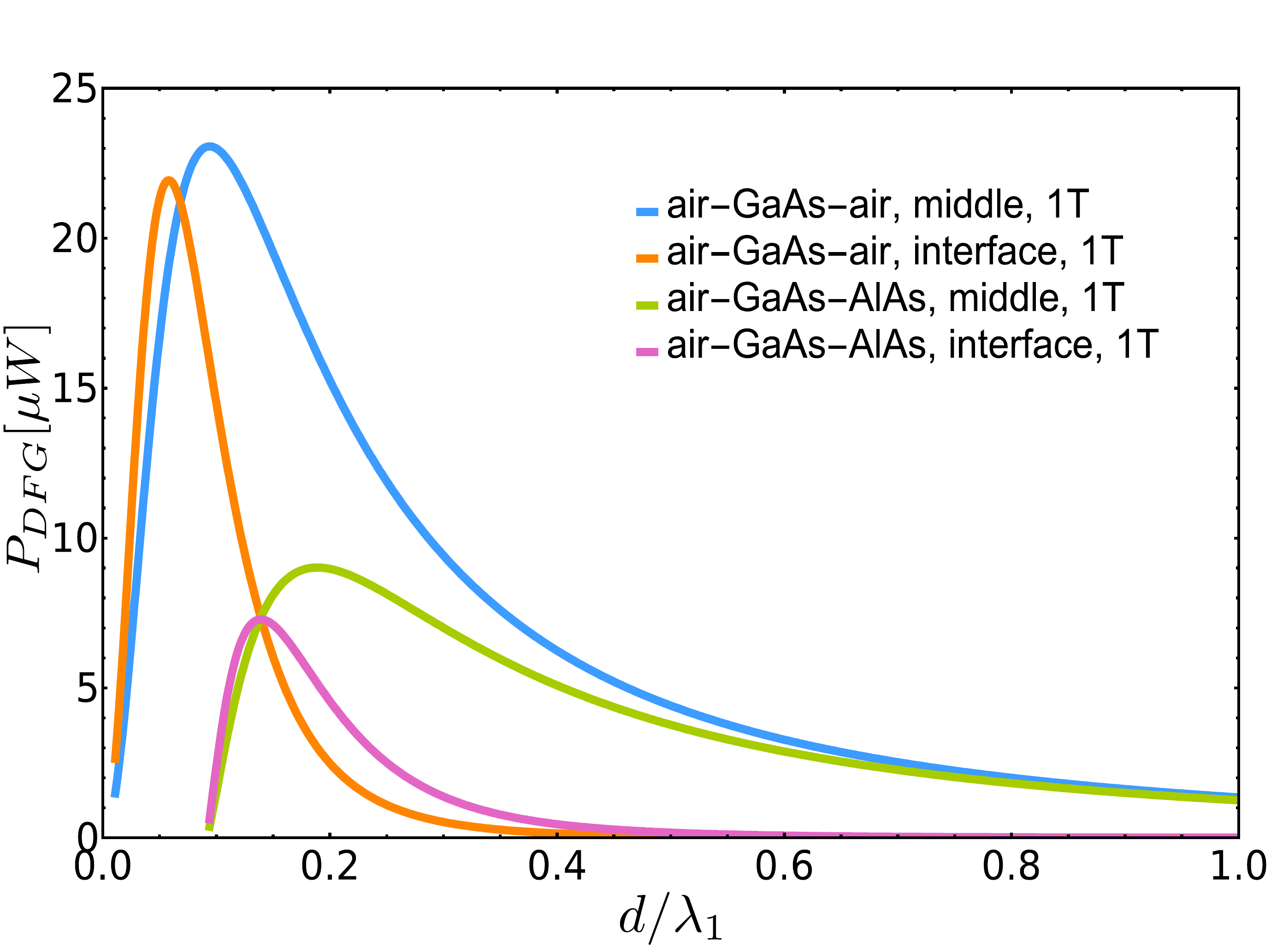}\label{automatedDFGpowerGaAS}
\caption{DFG power per 1 W$^2$ of the pump power as a function of the waveguide core thickness for the magnetic field strength 1T.  In the legend of the plot ``middle'' means that graphene is in the middle of the core dielectric $\epsilon_{2}$; ``interface'' means that graphene is located at the interface of dielectrics $\epsilon_{2}$ and $\epsilon_{3}$.  "1T" and "3T" stands for 1 and 3 Tesla magnetic field.}
\end{figure}

\begin{figure}[ht]
\includegraphics[clip,width=\columnwidth]{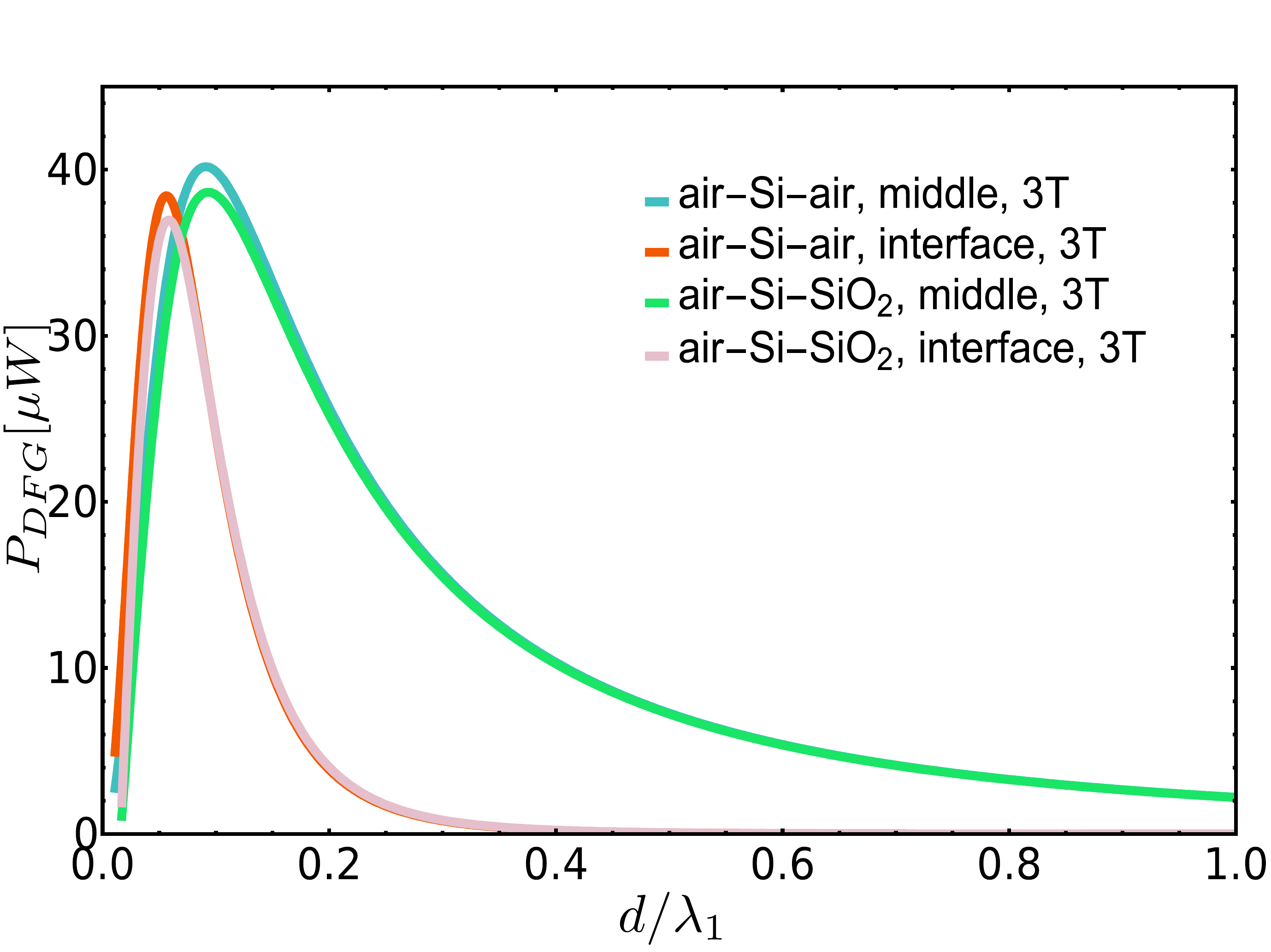}\label{automatedDFGpowerSi}
\caption{DFG power per 1 W$^2$ as a function of core thickness for the magnetic field strength 3 T. A higher magnetic field is chosen to avoid THz absorption in Si.}
\end{figure}

\begin{figure}[h]
    \centering
    \includegraphics[width=0.8\columnwidth]{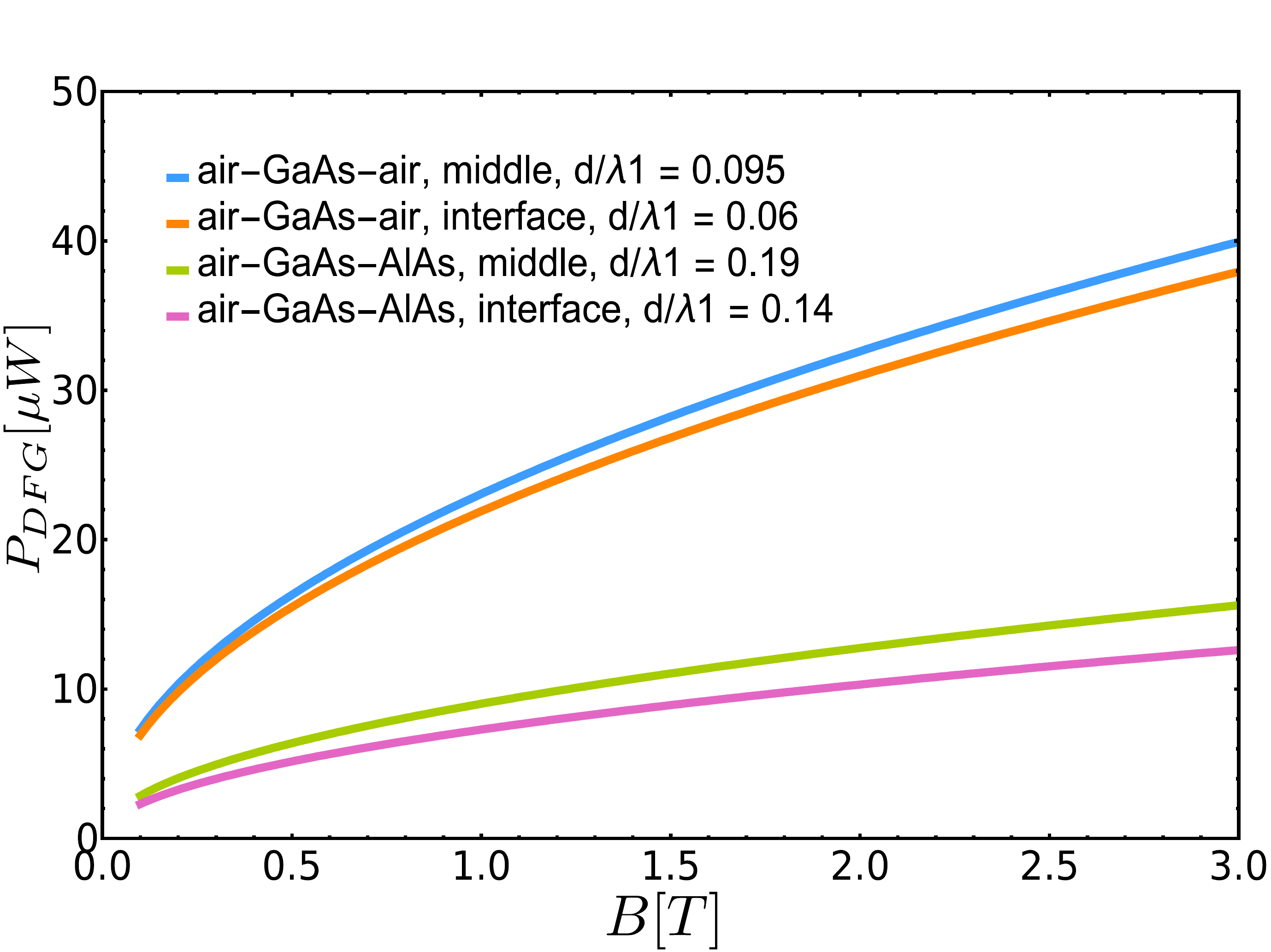}
    \caption{The DFG power per 1 W$^2$ as a function of the magnetic field for several waveguide structures and geometries. In the legend of the plot ``middle'' means that graphene is in the middle of the core dielectric $\epsilon_{2}$; ``interface'' means that graphene is located at the interface of dielectrics $\epsilon_{2}$ and $\epsilon_{3}$.}
	\label{fig6}
\end{figure}

\begin{figure}[h]
    \centering
    \includegraphics[width=0.8\columnwidth]{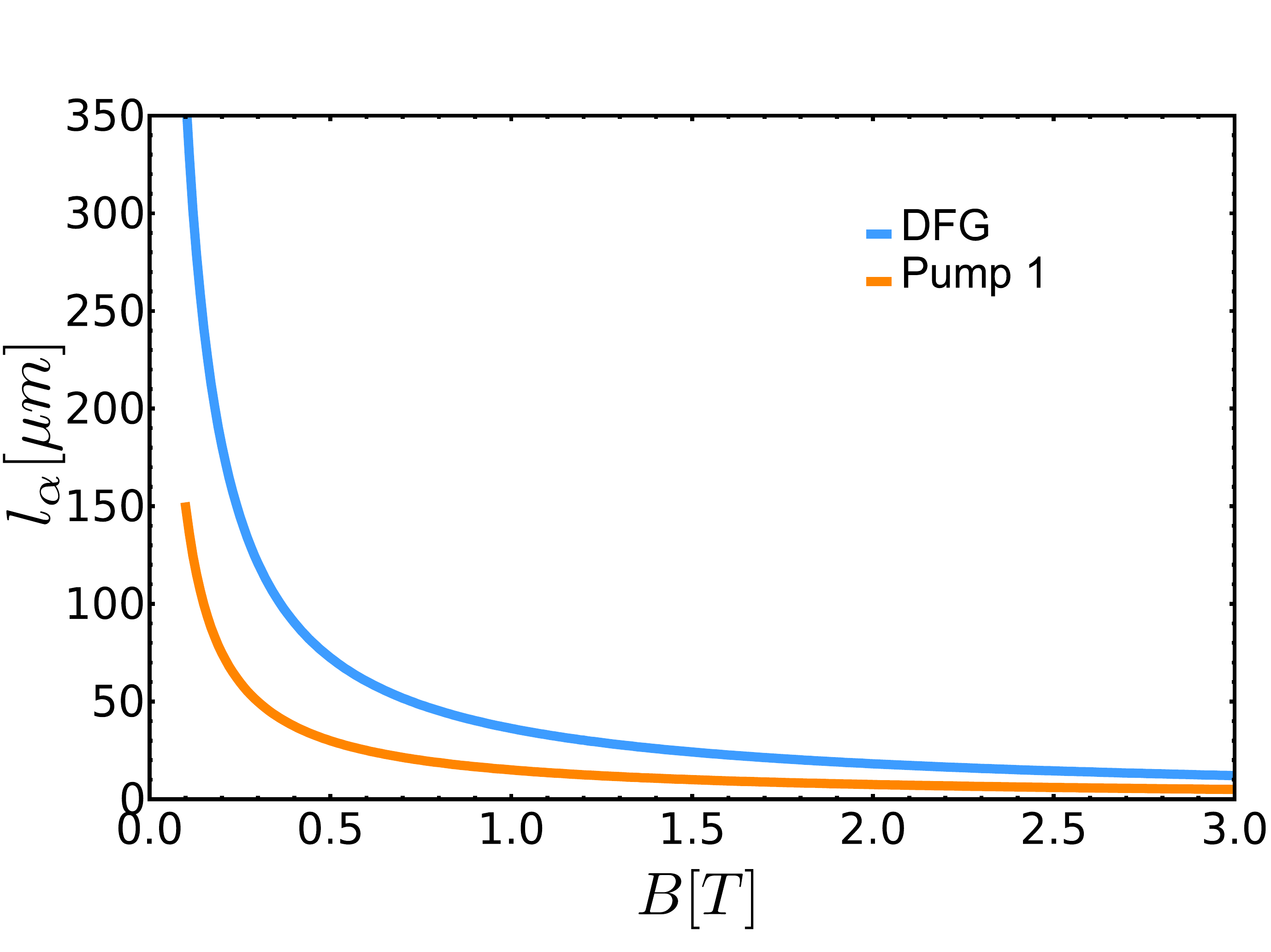}\label{absorptionlengthvsB}
    \caption{Absorption length for pump field intensity and DFG plasmon-polaritons as a function of the magnetic field for a symmetric GaAs waveguide with graphene at the interface.  The core thickness is $0.06\lambda_{1}$.}
	\label{fig8}
\end{figure}

\begin{figure}[h]
\includegraphics[clip,width=\columnwidth]{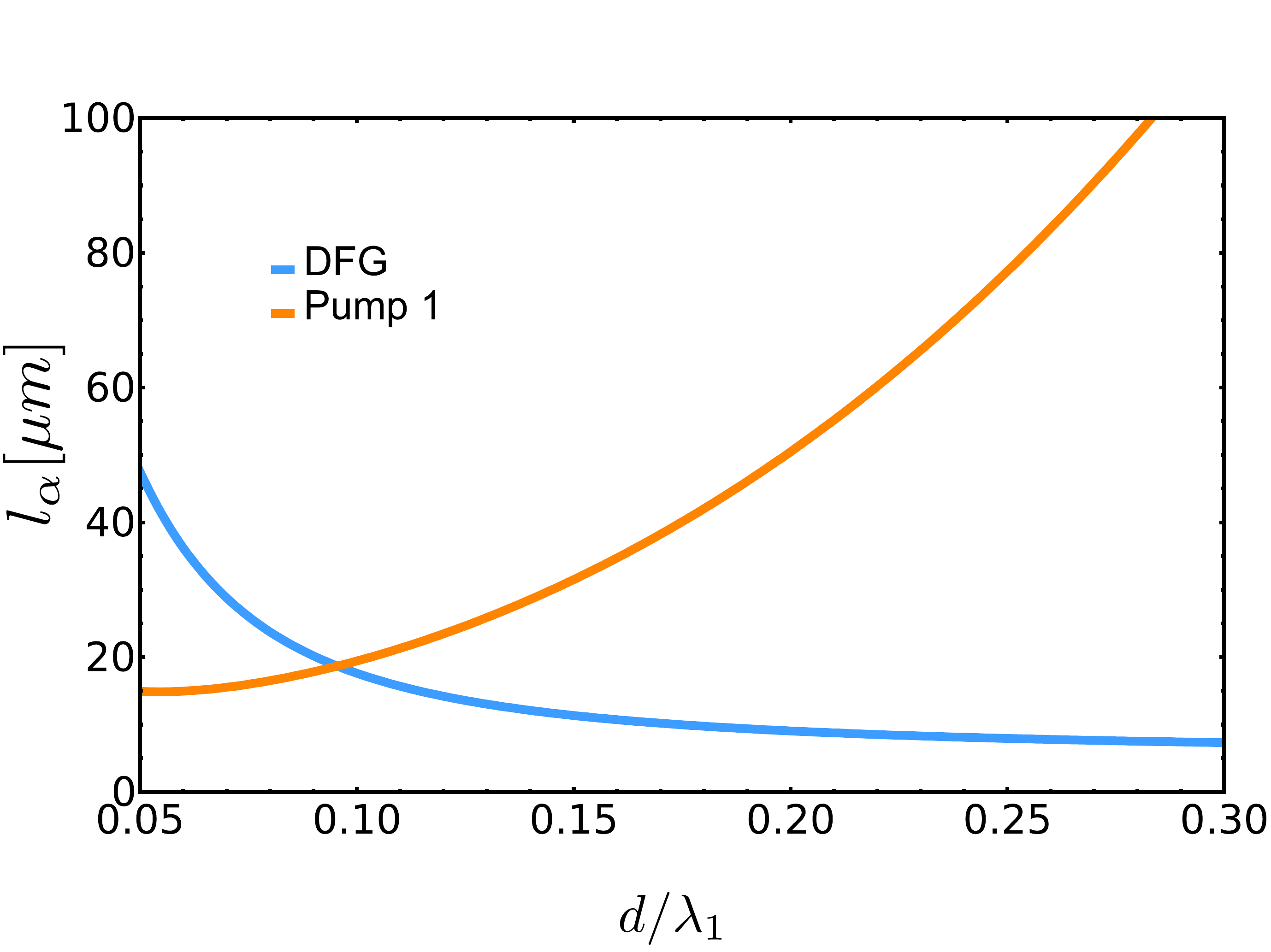}\label{absorptionlengthvst}
\caption{Absorption length for pump field intensity and DFG plasmon-polaritons as a function of core thickness for a symmetric GaAs waveguide with graphene at the interface.  The magnetic field is 1T.}
\label{fig7}
\end{figure}

 Figures 4-6 illustrate the dependence of the DFG power on various parameters for different waveguide compositions and locations of the graphene monolayer. The structure width $L_y$ is chosen to be 100 $\mu$m. The power scales linearly with $L_y$. For the plots we choose the initial state $\ket{1}$ in Fig.~2 to have the Landau level index $n=-3$. Then the states $\ket{2}$ and $\ket{3}$ coupled to state $\ket{1}$ by electric dipole-allowed pump transitions have Landau level numbers   $|n|-1=2$ and $|n|+1=4$, respectively. 
 The DFG frequency corresponding to the transition between these states is in the THz range; see Fig.~3. The pump wavelengths are in the mid-infrared; for example, at $B = 1$ T they are 10.9 $\mu$m and 9.1 $\mu$m. All frequencies scale as $\sqrt{B}$. The pump powers are assumed to be 1 W each, so that the plots actually show DFG power conversion efficiency in $\mu$W/W$^2$. 

 Figures 4 and 5 show the dependence of the DFG power on the thickness of the waveguide core for different positions of the graphene sheet and different waveguide materials at a fixed magnetic field. The DFG power depends on the magnitude of the in-plane components of the pump fields on graphene and the localization of the optical pump power. There is an optimal waveguide thickness which maximizes the DFG power for a given total power in the pump fields. For wider waveguide cores the in-plane component of the pump field amplitude on graphene gets smaller, whereas for narrower waveguides the pump field mode gets delocalized. Figures 4 and 5 also indicate that it is beneficial to place graphene in the middle of the waveguide core.  
 
 With increasing magnetic field the peak DFG power in Eq.~(\ref{DFGpower}) scales as  $\sqrt{B}$, provided the pump wavelengths are tuned in resonance with corresponding  transitions. This dependence is illustrated in Fig.~6 for a particular choice of waveguide structures and geometries. Note that the choice of particular pump and DFG transitions for a given magnetic field is strongly influenced by absorption in the waveguide materials. For example, one should obviously avoid reststrahlen bands in all waveguide layers. 
 
The DFG power can be further enhanced by stacking several monolayers together. However, there is  a trade-off between the nonlinear conversion efficiency and absorption in graphene. We calculated the absorption of both pump and difference frequency modes. 

The simplest way to calculate the absorption of the SPP mode  is to solve its dispersion equation Eq.~(\ref{altfulldispersion}) for a complex wavenumber $q$ as a function of a real frequency $\omega$, i.e. as a boundary-value problem. Then the absorption length of the plasmon field intensity is 
\begin{equation}
\label{abs-plasmon} 
l_{\rm abs} = \frac{1}{2} {\rm Im}q \simeq  \frac{1}{2} \gamma_{32} \left(\frac{\partial[\omega_o(q)]}{\partial q} \right)^{-1},
\end{equation} 
assuming $|{\rm Im}q| \ll |{\rm Re}q|$.

Among the two pump fields, the strongest absorption is experienced by the one  at frequency $\omega_{1}$  resonant with transition $\ket{1}\rightarrow\ket{3}$, because  state $\ket{1}$ is below the Fermi level whereas state $\ket{3}$ is above the Fermi level.  Its absorption length can be found from the linear conductivity calculated in Appendix D and the Poynting flux calculated in Appendix C:
\begin{equation}
\label{abs-pump} 
\frac{1}{l_{\rm abs}(\omega_1)} = \frac{1}{8} \frac{{\rm Re}\left[\sigma^{+-}(\omega_1) \right] \left|E_y^{(1)}(z = -d/2)\right|^2}{\expval{\Phi_S^{(1)}} }.   
\end{equation} 

The dependence of the absorption length from the magnetic field and the waveguide core thickness is shown in Figs.~7 and 8, assuming exact resonance with corresponding LL transitions and the linewidth of $10^{12}$ s$^{-1}$. This is a rather small linewidth corresponding to a high-quality graphene encapsulated in hBN. Therefore we probably overestimate the absorption rate for most samples and the actual absorption length is longer. In any case, for structures longer than the pump absorption length the pump field mode should be excited by a beam coupled from the top rather than from the facet, in order to reduce the propagation length. 

In conclusion, we investigated an electric-dipole-forbidden process of THz difference frequency generation in Landau-quantized graphene. The second-order susceptibility turned out to be surprisingly high, equivalent to the bulk magnitude of about $3\times 10^{-3}$ m/V. We applied the formalism to the DFG of THz surface plasmon-polaritons in graphene integrated into a dielectric waveguide or cavity  with strong vertical confinement of the optical pump modes. The DFG power conversion efficiency of the order of tens $\mu$W/W$^2$ is predicted from structures of size around $100$ $\mu$m. Analytic expressions for the DFG power are obtained and the results are presented for different structure geometries, composition, and magnetic field strengths. 

\section{acknowledgement} 

This material is based upon work supported by the Air Force Office of Scientific Research under award number  FA9550-17-1-0341.  M.T. acknowledges the support from RFBR grant No. 17-02-00387. 

\appendix

\section{ Eigenstates, optical matrix elements, and selection rules for Landau-quantized graphene}

For graphene in a constant external magnetic field $\bm{p}\rightarrow \bm{\pi}=\bm{p}+e\bm{A}/c$, where $\bm{p}$ is the canonical momentum, $\bm{\pi}$ is the gauge-invariant kinetic momentum and $\bm{A}$ is the vector potential that generates the magnetic field $\bm{B}=\nabla\times\bm{A}$.  The effective mass low-energy Hamiltonian (neglecting the spin degree of freedom) is then \cite{Goerbig2011}
\begin{equation}\label{maghamiltonian}
H_{\Xi}^{B}=\Xi v_{F}\bm{\sigma}\cdot\bm{\pi},
\end{equation} 
where $\bm{\sigma}$ is the vector of Pauli matrices and $\Xi = \pm 1$ depending on the valley. 
Assuming that the magnetic field $\bm{B}=\bm{e_z}B$ is perpendicular to the plane of the graphene sheet and using the Landau gauge $\bm{A}=-\bm{e_x}yB$, the eigenfunctions \cite{ZhengandAndo2002} are
\begin{equation}\label{wavfunK}
\psi^{K}_{nk}(\bm{r})=\frac{C_{n}}{\sqrt{L}}e^{ikx}
\mqty({\rm sgn}(n)i^{|n|-1}\phi_{|n|-1,k}(y) \\ i^{|n|}\phi_{|n|,k}(y) \\ 0 \\ 0) 
\end{equation}
and
\begin{equation}\label{wavfunK'}
\psi^{K'}_{nk}(\bm{r})=\frac{C_{n}}{\sqrt{L}}e^{ikx}
\mqty(0 \\ 0 \\ i^{|n|}\phi_{|n|,k}(y) \\ {\rm sgn}(n)i^{|n|-1}\phi_{|n|-1,k}(y)) 
\end{equation}
where
\begin{equation}\label{hermitepol}
\phi_{|n|,k}(y)=\frac{H_{|n|}((y-kl_{B}^{2})/l_{B})}{\sqrt{2^{|n|}|n|!}\sqrt{\pi}l_{B}}{\rm exp}\left[-\frac{1}{2}\left(\frac{y-kl_{B}^{2}}{l_{B}}\right)^{2}\right]
\end{equation}
with energy eigenvalue \cite{ZhengandAndo2002,McClure1956}
\begin{equation}\label{landauenergy}
\mathcal{E}_{n}= {\rm sgn}(n)\hbar\omega_{c}\sqrt{|n|};
\end{equation}
$L^2$ is the area of the system, $n=0,\pm 1, \pm 2,...$ is the Landau level index,  $\omega_{c}=\sqrt{2}v_F/l_B$ is the cyclotron frequency, $l_{B}=\sqrt{c\hbar/eB}$ is the magnetic length, $H_{|n|}((y-kl_{B}^{2})/l_{B})$ are the Hermite polynomials, $C_{n}=1$ for $n=0$ and $1/2$ otherwise.  Henceforth, all calculations will be carried out using the  effective mass low-energy Hamiltonian (\ref{maghamiltonian}) in the vicinity of the K point ($\Xi=+1$) and its eigenfunctions (\ref{wavfunK}).

The Hamiltonian for graphene in a magnetic field and an optical field is \cite{Yao2012,Booshehri2012}
\begin{equation}\label{tothamiltonian}
H=H^{B}+H^{opt} = v_{F}\bm{\sigma}\cdot\bm{\pi}+v_{F}\bm{\sigma}\cdot\frac{e\bm{A}^{opt}(t)}{c}
\end{equation}
$H^{opt}$ is the interaction Hamiltonian.

Note that the wavefunction (\ref{wavfunK}) can be written as $\psi_{n,k}(\bm{r})=\braket{\bm{r}}{n,k}$.  We'll make use of the state ket for graphene in a magnetic field $\ket{n,k}$ which will at times be written as $\ket{\alpha}$ for convenience.  

We utilize the notation above in calculating the optical matrix element for transitions between the LLs resonant with the optical field (in the electric dipole approximation)
\begin{equation}\label{opticalmel}
\mel{n,k'}{H^{opt}}{m,k}=v_{F}\frac{e\bm{A}^{opt}(t)}{c}\cdot\mel{n,k'}{\bm{\sigma}}{m,k}.
\end{equation}

  It is convenient to change to the circular polarization basis $\bm{e}_{\pm}\equiv 1/\sqrt{2}(\bm{e_x}\pm i\bm{e_y})$, termed left-circularly polarized (LCP) and right-circularly polarized (RCP), respectively.  The following relations holds true: $\bm{e}_{\pm}\cdot \bm{e}_{\pm}=0$ and $\bm{e}_{\pm}\cdot \bm{e}_{\mp}=1$.  In the $\bm{e}_{\pm}$ basis $Re[\bm{A}^{opt}]=c/i\omega_{opt}(\bm{e}_{+}E_{+}(t)+\bm{e}_{-}E_{-}(t))+c.c.$, where $E_{\pm}(t)=1/\sqrt{2}((E_{x}/2)\mp i(E_{y}/2))e^{-i\omega_{opt}t}$. 
 Similarly, the vector of Pauli matrices in the $\bm{e}_{\pm}$ basis is $\bm{\sigma}=\bm{e}_{+}\sigma^{+}+\bm{e}_{-}\sigma^{-}$  where 
\begin{equation} \label{A8} 
\sigma^{+}=\mqty(0 & 0 \\ \sqrt{2} & 0),\quad \sigma^{-}= \mqty(0 & \sqrt{2} \\ 0 & 0).
\end{equation}

   For a transition between Landau levels $m$ and $n$ resonant with the optical field ($\omega_{nm}\approx\omega_{opt}\equiv \omega$) we obtain
  
\begin{align}
&\mel{n,k'}{H^{opt}}{m,k}=\delta_{kk'}\sqrt{2}v_{F}C_{n}C_{m}(\bm{e}_{-}{\rm sgn}(n)\delta_{|n|-1,|m|}\nonumber\\
&+\bm{e}_{+}{\rm sgn}(m)\delta_{|n|+1,|m|})\cdot \left(e\frac{\bm{e}_{+}E_{+}(t)+\bm{e}_{-}E_{-}(t)}{i\omega}+c.c.\right)\nonumber\\
&\mel{n,k'}{H^{opt}}{m,k}=\delta_{kk'}\frac{\sqrt{2}v_{F}eC_{n}C_{m}}{i\omega}\times\nonumber\\
&({\rm sgn}(n)E_{+}(t)\delta_{|n|-1,|m|}+{\rm sgn}(m)E_{-}(t)\delta_{|n|+1,|m|})+c.c.\label{opticalmel2}
\end{align}
Equation (\ref{opticalmel2}) gives the selection rules for optical transitions between adjacent Landau levels i.e. $\Delta |m|=\pm 1$ \cite{Booshehri2012, Yao2013, Sadowski2006, Abergel2007}.  Furthermore, the transition $|m|\rightarrow|n|\pm1$ couples to the RCP/LCP component of the optical field, respectively.  From Eq.~(\ref{opticalmel2}) one also obtains the magnitude of the dipole moment \cite{Wang2015}
\begin{equation}\label{dipolemagnitude}
|d_{nm}|=\sqrt{2}C_{n}C_{m}\frac{ev_{F}}{\omega}.
\end{equation}

\section{Calculation of the matrix element $F_{nkmk'}(\bm{q})$ }

 Using the wavefunctions Eq.~(\ref{wavfunK}) with Eq.~(\ref{hermitepol}) the matrix element $F_{nkmk'}(\bm{q})$ can be calculated as
\begin{align}
&F_{nkmk'}(\bm{q})=\mel{n,k}{e^{iqx}}{m,k'}\label{Fmel1}\\
&F_{nkmk'}(\bm{q})=\frac{C_{n}C_{m}}{L}\int dx\,e^{i(k'-(k-q))x}\nonumber\\
&\times\int dy\,\left({\rm sgn}(n)i^{-|n|+1}\phi_{|n|-1,k}(y),i^{-|n|}\phi_{|n|,k}(y)\right)\nonumber\\
&\times \mqty({\rm sgn}(m)i^{|m|-1}\phi_{|m|-1,k'}(y) \\ i^{|m|}\phi_{|m|,k'}(y))\nonumber\\
&F_{nkmk'}=\frac{C_{n}C_{m}}{L}L\delta_{k',k-q}i^{|m|-|n|}\nonumber\\
&\times \int dy\,[{\rm sgn}(n){\rm sgn}(m)\phi_{|n|-1,k}(y)\times\phi_{|m|-1,k-q}(y)\nonumber\\
&+\phi_{|n|,k}(y)\times\phi_{|m|,k-q}(y)]\nonumber\\
&F_{nkmk'}=C_{n}C_{m}i^{|m|-|n|}\delta_{k',k-q}[{\rm sgn}(n) {\rm sgn}(m)\nonumber\\
&\times\braket{\phi_{|n|-1,k}}{\phi_{|m|-1,k-q}}+\braket{\phi_{|n|,k}}{\phi_{|m|,k-q}}]\label{Fmel2}.
\end{align}
Here
\begin{equation}
\int dx\,e^{i(k'-(k-q))x}=L\delta_{k',k-q}.
\end{equation}

Introducing the notation 
\begin{equation}\label{FmelTok}
F_{nkmk'}=\tilde{F}_{nkm}\delta_{k',k-q}
\end{equation}
and comparing equations (\ref{Fmel2}) and (\ref{FmelTok}) we see that
\begin{align}
\tilde{F}_{nkm}&=C_{n}C_{m}i^{|m|-|n|}[{\rm sgn}(n){\rm sgn}(m)\nonumber\\
&\times\braket{\phi_{|n|-1,k}}{\phi_{|m|-1,k-q}}+\braket{\phi_{|n|,k}}{\phi_{|m|,k-q}}].
\end{align}
We also have
\begin{equation}
{\rm sgn}(n){\rm sgn}(m)=
\begin{cases}
+1 &\quad \text{intraband transitions}\\
-1 &\quad \text{interband transitions}
\end{cases}.
\end{equation}
In the main text we have the states labeled in the following way: $\ket{1}=\ket{-|m|}$, $\ket{2}=\ket{|m|-1}$, $\ket{3}=\ket{|m|+1}$.  With this labeling, the  second-order nonlinear susceptibility and the corresponding SPP field contain the matrix element  $F_{3k2k'}=\tilde{F}_{3k2}\delta_{k',k-q}$.  So for the initial state of $m\neq 0$ we have 
\begin{align}
&F_{3k2k'}\rightarrow\nonumber\\
&F_{|m|+1,k,|m|-1,k'}=\delta_{k',k-q}C_{|m|+1}C_{|m|-1}i^{|m|-1-(|m|+1)}\nonumber\\
&\times\left[\braket{\phi_{|m|+1-1,k}}{\phi_{|m|-1-1,k-q}}+\braket{\phi_{|m|+1,k}}{\phi_{|m|-1,k-q}}\right]\nonumber\\
&F_{3k2k'}\rightarrow F_{|m|+1,k,|m|-1,k'}=-\delta_{k',k-q}C_{|m|+1}C_{|m|-1}\nonumber\\
&\times\left[\braket{\phi_{|m|,k}}{\phi_{|m|-2,k-q}}+\braket{\phi_{|m|+1,k}}{\phi_{|m|-1,k-q}}\right]
\end{align}
For the initial state $m=-3$ we have
\begin{align}
&F_{|m|+1,k,|m|-1,k'}\rightarrow\nonumber\\
&F_{4k2k'}=-\frac{1}{2}\delta_{k',k-q}\left[\braket{\phi_{3,k}}{\phi_{1,k-q}}+\braket{\phi_{4,k}}{\phi_{2,k-q}}\right]\nonumber\\
&\tilde{F}_{4k2}=-\frac{1}{2}\left[\braket{\phi_{3,k}}{\phi_{1,k-q}}+\braket{\phi_{4,k}}{\phi_{2,k-q}}\right]
\end{align}
The analytic expression for $\tilde{F}_{4k2}$ is 
\begin{align}
&\tilde{F}_{4k2}(q)=-\left[\frac{24 (2 + \sqrt{2}) - 4 (4 + \sqrt{2}) l_{B}^{2} q^{2} + l_{B}^{4} q^{4}}{128\sqrt{3}}\right]\nonumber\\
&\times l_{B}^{2}q^{2}e^{-(l_{B}^{2}q^{2}/4)}\label{Fmel4k2}\\
&\tilde{F}_{4k2}(q)\approx -\left[\frac{24 (2 + \sqrt{2})}{128\sqrt{3}}\right]l_{B}^{2}q^{2}
\end{align}
The factor $\zeta(q)$ in the main text is defined as follows:
\begin{equation}
\zeta(q)=\frac{1}{\kappa}\sum_{k'}\tilde{F}_{2k'4}(-q)\label{zetaeqn}
\end{equation}
where $\kappa = 2S/\pi l_{B}^{2}$ and $l_{B} = \sqrt{c\hbar/eB}$.  From Eq.~(\ref{Fmel4k2}) we see that $\tilde{F}_{2k'4}(-q)$ is independent of $k'$ and can be taken out of the sum.  Next we use:
\begin{equation}
\sum_{k'}\to 4L_{x}L_{y}/2\pi l_{B}^{2} = \kappa \label{sumk}
\end{equation}
using Eq.~(\ref{sumk}) in Eq.~(\ref{zetaeqn}) we find
\begin{equation}
\zeta(q)=\tilde{F}^{*}_{4k'2}(q)\propto l_B^{2}\propto 1/B.
\end{equation}

\section{ Normalization of pump fields}
We begin by considering the waveguide structure where the interfaces are at $z=d/2$ and $z=-d/2$.  The thickness of the core layer is $d$ and a monolayer of graphene is located at the interface $z=-d/2$.  The dielectric constant is then:
\begin{equation}\label{epszvar}
\epsilon_{j}=
\begin{cases}
\epsilon_{1} & \quad z>d/2\\
\epsilon_{2} & \quad -d/2<z<d/2\\
\epsilon_{3} & \quad z<-d/2.
\end{cases}
\end{equation}
We have two counter propagating TE polarized pump fields in the waveguide $\bm{E}^{1,2}(x,z,t)= {\rm Re}[(0,E_{y}^{1,2}(z),0)e^{\pm iq_{1,2}x-i\omega_{1,2} t}]$.  

Both pump fields obey the wave equation (in each region of the waveguide indexed by $j$):
\begin{align}
&(\nabla^{2}-\frac{\epsilon_j}{c^2}\partial^{2}_{t})\bm{E}^{l}_{j}(x,z,t)=0\\
\implies & \frac{d^{2}E^{l}_{jy}(z)}{dz^{2}}=\lambda_{jl}E^{l}_{jy}(z)\label{waveqgen}.
\end{align}
Here the eigenvalue determining the confinement of the pump field to the core layer of the waveguide is
\begin{equation}\label{eigenvalsacrossz}
\lambda_{jl}=
\begin{cases}
+\kappa^{2}_{1l} & \quad z>d/2\\
-\alpha^{2}_{l} & \quad -d/2<z<d/2\\
+\kappa^{2}_{3l} & \quad z<-d/2
\end{cases}
\end{equation}
where 
\begin{align}
\kappa_{(1,3)l}=&\sqrt{q^{2}_{l}-\epsilon_{1,3}\frac{\omega^{2}_{l}}{c^{2}}}\\
\alpha_{l}=&\sqrt{\epsilon_{2}\frac{\omega^{2}_{l}}{c^{2}}-q^{2}_{l}}
\end{align}
with the confinement condition $n_{1,3}<n_{eff}<n_{2}$ where $n_{j}^{2}\approx \epsilon_{j}$ for small losses.  The solution to the eigenvalue equation (\ref{waveqgen}) along with the continuity of the tangential component $E_{jy}(z)$ of the pump fields at interfaces $z=\pm d/2$ gives:
\begin{equation}\label{wavgeneqsol}
E_{jy}^{l}(z)=A_{l}f_{jl}(z)
\end{equation}
\begin{equation}\label{fieldprofile}
f_{jl}(z)=
\begin{cases}
\cos(\alpha_{l}d/2-\phi_{l})e^{-\kappa_{1l}(z-d/2)} & \; z>d/2\\
\cos(\alpha_{l}z-\phi_{l}) & \; -d/2<z<d/2\\
(\cos(\alpha_{l}d/2+\phi_{l})e^{\kappa_{3l}(z+d/2)} & \; z<-d/2 
\end{cases}
\end{equation}
We will drop the index $j$ for derivations that follow while keeping in mind that the field profile $f_{jl}(z)$ is a piecewise function.  Next we find the amplitude $A$ by normalizing the average Poynting flux $\expval{\Phi_{S}}$ to 1 W.  We will drop the superscript $l$  of the fields. 

Let $\bm{F}$ represent $\bm{E}$ and $\bm{B}$.  We can write the field $\bm{F}$ as $\bm{F}=Re[\bm{F}_{R}e^{-i\omega t}]=\frac{1}{2}(\bm{F}_{R}e^{-i\omega t}+\bm{F}_{R}^{*}e^{i\omega t})$, where $\bm{F}_{R}=\bm{F}_{o}e^{i\bm{k}\cdot\bm{r}}$.  For a TE polarized field $\bm{E}_{R}=(0,E_{y}(z),0)e^{iqx}$, $\bm{B}_{R}=(B_{x}(z),0,B_{z}(z))e^{iq x}$.  
The time average of the Poynting flux is:
\begin{equation}\label{avepoyntingflux}
\expval{\Phi_{S}}=\frac{c}{8\pi}\int d\bm{A}\cdot Re[\bm{E}_{R}\times\bm{B}_{R}^{*}]
\end{equation}
or, for the pumps propagating along the x-direction,  
\begin{equation}
\expval{\Phi_{S}}=\frac{c}{8\pi}\int dydz\, Re[E_{y}(x,z)B_{z}^{*}(x,z)]
\end{equation}
We will assume that the fields are uniform along $y$, so integration over $y$ results in multiplying by the length of the waveguide in the y-direction, $L_{y}$. 

From Maxwell equations  for a TE mode
\begin{align}
&B^{*}_{z}(x,z)=\frac{cq}{\omega}E^{*}_{y}(x,z).
\end{align}
The Poynting flux for TE pump fields is therefore
\begin{equation}\label{gennormalizedpoyntingflux}
\expval{\Phi^{l}_{S}}=\frac{q_{l} L_{y}c^{2}}{8\pi\omega_{l}}\int_{-\infty}^{\infty} dz\,|E^{l}_{y}(z)|^{2}
\end{equation}
Finally, we normalize the Poynting flux $\expval{\Phi_{S}}$ as $\expval{\Phi_{S}}=P_{o}$, where $P_{o}$ is the input pump power. This gives  
\begin{align}\label{normalizedfieldamplitudes}
&A_{l}=\sqrt{\frac{P_{o}}{\frac{q_{l} L_{y}c^{2}}{8\pi\omega_{l}}F_{l}}}
\end{align}
where 
\begin{equation}
F_{l}=\int_{-\infty}^{\infty} dz\,|f_{l}(z)|^{2}.
\end{equation}

\section{ Linear conductivity of Landau-quantized graphene}

Linear conductivity of Landau-quantized graphene has been calculated a number of times before. Here we summarize one approach to the derivation,  which is based on  the density matrix equation where the Hamiltonian is given by Eq.~(\ref{tothamiltonian}),   
\begin{align}
\dot{\rho}_{\alpha\beta}&=\frac{i}{\hbar}\left[\rho, H\right]_{\alpha\beta}-\gamma_{\alpha\beta}\rho_{\alpha\beta}\nonumber \\
&=\frac{i}{\hbar}(H_{\beta\beta}-H_{\alpha\alpha})\rho_{\alpha\beta}+\frac{i}{\hbar}(\rho_{\alpha\alpha}-\rho_{\beta\beta})H_{\alpha\beta}-\gamma_{\alpha\beta}\rho_{\alpha\beta}\label{densmateq3}
\end{align}
$\rho_{\alpha\beta}$ is the density matrix element, $\gamma_{\alpha\beta}$ is the phenomenological decay term.  The states $\ket{\alpha}$ are eigen-states of the Hamiltonian $H^{B}$, i.e. $H^{B}\ket{\alpha}=\mathcal{E}_{\alpha}\ket{\alpha}$.  In the dipole approximation (\ref{densmateq3}) becomes:
\begin{align}
&\dot{\rho}_{\alpha\beta}(t)=i(-\omega_{\alpha\beta}+i\gamma_{\alpha\beta})\rho_{\alpha\beta}(t)-\frac{1}{i\hbar}(f_{\alpha}-f_{\beta})\nonumber\\
&\times ev_{F}\bm{\sigma}_{\alpha\beta}\cdot\left(\frac{\bm{E}}{i\omega_{l}}e^{-i\omega_{l} t}+c.c.\right)\label{densmateq5}
\end{align}
where $f_{\alpha}=\rho_{\alpha\alpha}$ is $1$ if the state $\ket{\alpha}$ is occupied or $0$ if it's unoccupied; $\omega_{\alpha\beta}=(\mathcal{E}_{\alpha}-\mathcal{E}_{\beta})/\hbar$, $\bm{E}=\bm{e}_{+}E_{+}+\bm{e}_{-}E_{-}$.\newline{}
In the rotating wave approximation
\begin{align}
\rho_{\alpha\beta}(t)&=\frac{i(f_{\alpha}-f_{\beta})ev_{F}\bm{\sigma}_{\alpha\beta}\cdot\bm{E}}{\hbar\omega_{l}(\omega_{l}-\omega_{\alpha\beta}+i\gamma_{\alpha\beta})}e^{-i\omega_{l}t}\\
&\equiv\rho_{\alpha\beta}(\omega_{l})e^{-i\omega_{l}t}\label{rhosol1}.
\end{align}
Note that the term
\begin{equation}
\frac{ev_{F}}{i\omega_{l}}\bm{\sigma}_{\alpha\beta}\cdot\bm{E}e^{-i\omega_{l}t}=\mel{\alpha}{H^{opt}}{\beta}; \nonumber
\end{equation}
the equality holds when we drop the complex conjugate part of $H^{opt}$.  The right-hand side of the equation above was calculated in Eq.~(\ref{opticalmel2}).  We extract the following terms from Eq.~(\ref{opticalmel2}) for the Pauli matrix elements defined in Eq.~(\ref{A8}):
\begin{align}
&\sigma^{+}_{\alpha\beta}=\mel{\alpha}{\sigma^{+}}{\beta}=C_{n}C_{m}\delta_{k,k'}\sqrt{2}{\rm sgn}(m)\delta_{|n|,|m|-1}\label{melsigma+alphabeta}\\
&\sigma^{-}_{\alpha\beta}=\mel{\alpha}{\sigma^{-}}{\beta}=C_{n}C_{m}\delta_{k,k'}\sqrt{2}{\rm sgn}(n)\delta_{|n|-1,|m|}\label{melsigma-alphabeta}.
\end{align}

The optical conductivity of graphene can be obtained from the expectation value of the 2D current density $\expval{\bm{j}(t)}$.  
\begin{align}
&\expval{\bm{j}(t)}=tr(\rho(-\frac{e}{S}\bm{v}))=-\frac{e}{S}\sum_{\alpha}\sum_{\beta}\rho_{\alpha\beta}(t)\bm{v}_{\beta\alpha}\nonumber\\
&=-\frac{e}{S}\sum_{\alpha}\sum_{\beta}\rho_{\alpha\beta}(t)v_{F}\bm{\sigma}_{\beta\alpha}\\
&\expval{\bm{j}(t)}=-i\frac{e^{2}v_{F}^{2}}{S\hbar\omega_{l}}\sum_{\alpha}\sum_{\beta}\frac{(f_{\alpha}-f_{\beta})(\bm{\sigma}_{\alpha\beta}\cdot\bm{E})\bm{\sigma}_{\beta\alpha}}{(\omega_{l}-\omega_{\alpha\beta}+i\gamma_{\alpha\beta})}e^{-i\omega_{l}t}\\
&=\expval{\bm{j}(\omega_{l})e^{-i\omega_{l}t}}\label{current}.
\end{align}
In the component form
\begin{align}
\expval{j^{p}(\omega_{l})}&=-i\frac{e^{2}v_{F}^{2}}{S\hbar\omega_{l}}\sum_{r \neq p}\sum_{\alpha}\sum_{\beta}\frac{(f_{\alpha}-f_{\beta})\sigma^{p}_{\beta\alpha}\sigma^{r}_{\alpha\beta}}{(\omega_{l}-\omega_{\alpha\beta}+i\gamma_{\alpha\beta})}E_{p}\\
&\equiv\sum_{r \neq p}\sigma^{pr}_{con}(\omega_{l})E_{p},
\end{align}
where the indices $p$ and $r$ span over $+,-$. 

There are four components of the conductivity tensor that we need to calculate: $\sigma^{++}_{con}$, $\sigma^{--}_{con}$, $\sigma^{+-}_{con}$, and $\sigma^{-+}_{con}$. The first two of them are equal to zero. The only nonzero elements are 
\begin{align}
&\sigma^{+-}_{con}(\omega_{l})=-i\frac{e^{2}v_{F}^{2}}{S\hbar\omega_{l}}\sum_{\alpha}\sum_{\beta}\frac{(f_{\alpha}-f_{\beta})\sigma^{+}_{\beta\alpha}\sigma^{-}_{\alpha\beta}}{(\omega_{l}-\omega_{\alpha\beta}+i\gamma_{\alpha\beta})}\nonumber \\
&=-\frac{ie^{2}v_{F}^{2}}{\hbar\omega_{l}}\frac{1}{\pi l_{B}^{2}}\sum_{m}\frac{f_{|m|+1}-f_{m}}{\omega_{l}-\omega_{|m|+1,m}+i\gamma_{|m|+1,m}}\label{sigma+-reduced}.
\end{align}

\begin{align}
&\sigma^{-+}_{con}(\omega_{l})=-i\frac{e^{2}v_{F}^{2}}{S\hbar\omega_{l}}\sum_{\alpha}\sum_{\beta}\frac{(f_{\alpha}-f_{\beta})\sigma^{-}_{\beta\alpha}\sigma^{+}_{\alpha\beta}}{(\omega_{l}-\omega_{\alpha\beta}+i\gamma_{\alpha\beta})}\nonumber \\
& =-\frac{ie^{2}v_{F}^{2}}{\hbar\omega_{l}}\frac{1}{\pi l_{B}^{2}}\sum_{m}\frac{f_{|m|-1}-f_{m}}{\omega_{l}-\omega_{|m|-1,m}+i\gamma_{|m|-1,m}}\label{sigma-+reduced}
\end{align}
 $\sigma^{+-}_{con}$ couples to the $E_{+}$ component of $\bm{E}$ and $\sigma^{-+}_{con}$ couples to the $E_{-}$ component of $\bm{E}$.  The TE-polarized pump fields in our problem are $\bm{E}^{1,2}=(\bm{E}^{(1,2)}_{+}+\bm{E}^{(1,2)}_{-})e^{-i\omega_{1,2}t}+c.c.$.  We also have $\omega_{|m|+1,-|m|}$ resonant with $\omega_{1}$ and $\omega_{|m|-1,-|m|}$ resonant with $\omega_{2}$.  For definiteness let's assume the initial state is $m=-3$ so $\omega_{1}$ is resonant with $\omega_{4,-3}$ and $\omega_{2}$ with $\omega_{2,-3}$.  If we select only the resonant frequency then the conductivity becomes: 
\begin{align}
&\sigma^{+-}_{con}(\omega_{1})=-\frac{ie^{2}v_{F}^{2}}{\hbar\omega_{1}}\frac{1}{\pi l_{B}^{2}}\frac{f_{4}-f_{-3}}{\omega_{1}-\omega_{4,-3}+i\gamma_{4,-3}}
\end{align}
where we expect $f_{4}=0$ and $f_{-3}=1$.  Note that $\sigma^{+-}(\omega_{1})$ couples to $E^{(1)}_{+}$.\newline{}
Similarly,
\begin{align}
&\sigma^{-+}_{con}(\omega_{2})=-\frac{ie^{2}v_{F}^{2}}{\hbar\omega_{2}}\frac{1}{\pi l_{B}^{2}}\frac{f_{2}-f_{-3}}{\omega_{2}-\omega_{2,-3}+i\gamma_{2,-3}}  
\label{cond-+resonant},
\end{align}
where we expect $f_{2}-f_{-3} \simeq 0$ since both states are below the Fermi level.

\end{document}